\documentclass[9pt,twocolumn,twoside]{osajnl}

\journal{josab} 

\setboolean{shortarticle}{false}

\title{Quantum Metamaterials: Applications in quantum information science.}

\author[1,*]{SOLOMON URIRI}
\author[1]{YASEERA ISMAIL}
\author[1,2,3]{FRANCESCO PETRUCCIONE}

\affil[1]{Quantum Research Group, School of Chemistry and Physics, University of KwaZulu-Natal, University Road, Durban, 4000, South Africa}
\affil[2]{National Institute for Theoretical Physics, South Africa}
\affil[3]{School of Electrical Engineering, KAIST, Daejeon, 34141, Republic of Korea}

\affil[*]{Corresponding author: uriris@ukzn.ac.za}

\begin{abstract}
Metamaterials are artificially engineered periodic structures with exceptional optical properties that are not found in conventional materials.  However, this definition of metamaterials can be extended if we introduce a quantum degree of freedom by adding some quantum elements (e.g quantum dots, cold atoms, Josephson junctions, molecules). Quantum metamaterials can then be defined as artificially engineered nanostructures made up of quantum elements.  Furthermore, they exhibit controllable quantum states, maintain quantum coherence for times much higher than the transversal time of the electromagnetic signal. Metamaterials have been used to realised invisibility cloaking, super-resolution, energy harvesting, and sensing. Most of these applications are performed in the classical regime. Of recent, metamaterials have gradually found their way into the quantum regime, particularly to quantum sensing and quantum information processing. The use of quantum metamaterials for quantum information processing is still new and rapidly growing. In quantum information processing, quantum metamaterials have enabled the control and manipulation of quantum states, single photon generation, creating quantum entanglement, quantum states switching, quantum search algorithm, quantum state engineering tasks, and many more. In this work, we briefly review the theory, fabrication and applications of quantum metamaterials to quantum information processing. 
\end{abstract}

\setboolean{displaycopyright}{true}

\begin{document}

\maketitle

\section{Introduction}
Metamaterials are artificially engineered materials which have exotic optical properties that are not found in conventional materials. Metamaterials research was started in 1967 by Vaselego \cite{veselago1968}, who theoretically predicted materials with simultaneously effective negative permeability $\epsilon$ and effective magnetic permeability $\mu$, which give a negative refractive index. However, metamaterials began to attract more attention in the scientific community in 1997, when Pendry \cite{Pendry2000} demonstrated the perfect lens and took a major leap in 2001 when Smith et. al \cite{Smith2000} experimentally realized the simultaneous effective negative $\epsilon$ and $\mu$ for the first time. This was a breakthrough, leading to some important and useful applications which include electromagnetic cloak of a physical object using transformation optics \cite{AluEngheta2008} and super-resolution imaging \cite{Kim2010}. Traditionally, metamaterials are made from metallic nanostructures with size much smaller than the incident wavelength. The metallic nanostructures are arranged periodically, with the collective behaviour of the nanostructures. This gives rise to the bulk optical response of the material.  
\paragraph{} Metamaterials are classified into various forms according to the material constituents, such as hyperbolic metamaterials \cite{Cortes2012}, dielectric metamaterials \cite{zhang2018}, negative-index metamaterials \cite{Quach2011}, plasmonic metamaterials \cite{uriri2018}, superconducting metamaterials \cite{mirhosseini2018}, chiral metamaterials \cite{andrews2018}, and recently, quantum metamaterials \cite{Asano2015}. In optics, most of the studies involving metamaterials fall within the classical regime. However, little is currently known on the applications of quantum metamaterials, particularly to the areas of quantum communication and quantum computing. 
\begin{figure}
  \includegraphics[width=\linewidth]{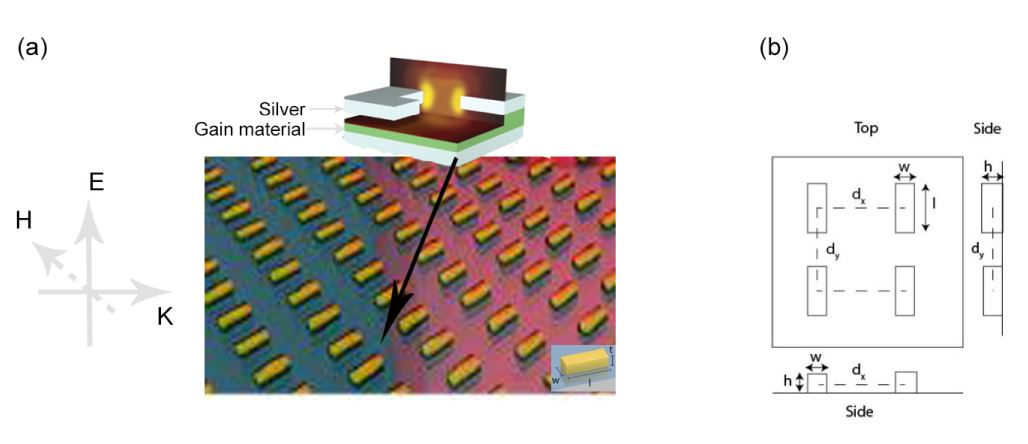}
  \caption{A schematic of a quantum plasmonic metamaterial with a gain media (a) and a schematic of the nanorods with dimensions $d_x$ = $d_y$ = 200 nm, t = 30 nm, l = 110 nm, and w = 45 nm (b). Here, $d_x$ = $d_y$ is the period, t = thickness, l = length, and = width. }
  \label{metamaterial}
\end{figure}
Quantum metamaterials (QMM) are artificially engineered material whose optical properties are determined by the interplay of quantum effects in the constituent atoms with the electromagnetic field modes in the system. Quantum metamaterials are hence an artificial optical material within which the individual unit cells demonstrate quantum coherence. Since nanostructures unit cells can preserve quantum coherence, QMM, are being realized as having important applications in quantum communication. Interest in the study of the applications of quantum metamaterials to quantum communication is therefore growing rapidly and has yielded some remarkable success. For example, quantum entanglement, a unique quantum resource for quantum communication, can easily be realized in QMM, where in Ref. \cite{Asano2015}, the authors used a plasmonic metamaterial to carry out the distillation of quantum entanglement. Other applications of QMM are found in the making of polarizers \cite{zhao2012}, half and quarter waveplates \cite{nouman2016}, reconfigurable quantum  superlenses \cite{Quach2011}, as well as for the implementation of quantum algorithms \cite{zhang2018}, performing quantum state engineering \cite{uriri2018}, and quantum state polarisation control \cite{Wang2018}. In this work, we briefly review the theory, fabrication, applications, challenges and future direction of quantum metamaterials related to quantum information processing. 

\section{Theory}
Localised surface plasmons (LSPs) are non-propagating excitation of the free electrons of metallic nanoparticles coupled to the electromagnetic field. LSPs are excited when the light of a particular frequency is incident on a metal nanoparticles of sub-wavelength size.  A conventional metamaterial can be described by the material's effective macroscopic parameters, which include $\epsilon$ and $\mu$. These parameters $\epsilon$ and $\mu$ are easily described by the Maxwell's equations. However, from the microscopic point of view, these parameters are functions of the average quantum states of the individual building blocks \cite{Quach2011}. Interestingly, these states can be directly controlled and maintain phase coherence over the relevant spatial and temporal scales in quantum metamaterials. Microscopically, the propagation of single photon states or qubits (that is, a two-level quantum system) through the quantum metamaterials can be fully described using the Jaynes-Cummings-Hubbard Hamiltonian defined as \cite{Quach2011}
\begin{equation}\label{one}
\mathcal{H} = \mathcal{H}_0 + \mathcal{H}_{QMM} + \mathcal{H}_{int},
\end{equation}
and written explicitly as
\begin{equation}
\mathcal{H} = \sum_j\hbar \omega_j b^{\dagger}_j b_j - \frac{1}{2}\sum_k(\epsilon_k \sigma_z^{(k)}+\Delta_k \sigma_x^{(k)}) + i \sum_{jk} \epsilon_{jk}\sigma_z^{(k)}(b_j - b_j^{\dagger}),
\end{equation}
where the first term represents the unperturbed photon mode, the second term is the qubit degrees of freedom and the third term describes the interaction of the photon mode through the quantum metamaterial. Using the quantum metamaterial, the control and the manipulation of the qubits is obtained via the qubit Hamiltonian parameter. Note this is acquired through the matrix element $\Delta$ or via $\epsilon$. Such an interaction form a photon wave packet with characteristic size $\Delta \gg s$, and averaging over the quantum states of qubits or single photons on the scale of $\wedge$. $\wedge$ describes the state of both the electromagnetic field and the quantum metamaterial, characterised by a non-diagonal, nonlocal, position- and state-dependent effective refractive index. Here, $s$ is the size of the nanostructure unit cells. However, the above equation is mostly applied to superconducting quantum metamaterials. For optical quantum metamaterials, some new quantum effects are added to the system Hamiltonian given Eq. \ref{one}, whereby the individual terms correspond to
\begin{equation}
\mathcal{H}_0 = \hbar\omega_0\hat{a}^{\dagger}\hat{a}+\hbar\omega_x\hat{\sigma}^{\dagger}\hat{\sigma}
\end{equation}
\begin{equation}
H_{QMM} = - E_0 \mu (\hat{\sigma} e^{-i\omega t} + \hat{\sigma}^\dagger e^{+ i\omega t}) - E_0 \chi (\hat{a} e^{- i\omega t} + \hat{a}^\dagger e^{i\omega t})
\end{equation}
\begin{equation}
\mathcal{H}_{int}=i\hbar g(\hat{a}\hat{a}^{\dagger} - \hat{\sigma}^{\dagger}\hat{\sigma}), 
\end{equation}
where $\omega_0$ and $\omega_x$ are the resonance frequencies of the metallic nanoparticle plasmonic field mode and the quantum dot. $\hat{a}^{\dagger}\hat{a}$ and $\hat{\sigma}^{\dagger}\hat{\sigma}$ are the creator and annihilation operators for the metallic nanoparticles plasmonic field mode and the quantum dot, respectively. In this last model, a quantum dot was added to the plasmonic metamaterial to increase the emission rate in the material. $\omega_0$ can be derived by using the Frohlish condition, and the derivation details are given in \cite{mcenery2014}.

An analytical model for active metamaterial with quantum elements was proposed by Chipouline et. al \cite{Chipouline2012} to describe the complex dynamics of a hybrid system consisting of reasonably coupled classical resonators and quantum structures. The model of Chipouline and his colleagues uses a density matrix approach together with an ideal harmonic oscillator equation to describe (for modelling) subwavelength plasmonics and optical resonators. In 2012, Zhou et. al \cite{Zhou2012} theoretically characterized a negative index metamaterial slab using pairs of entangled photons and a modified Hong-Ou-Interferometer. The authors reported the properties of dispersion and absorption of a negative-index metamaterial slab confirmed in fourth-order interference profile whereby a comparison with a coherent light shows the advantages with a quantum field. Kamli et. al \cite {Kamli2008} proposed fast all-optical control of surface polaritons by placing an electromagnetically induced transparency medium at an interface between two nanomaterials using a negative-index material. A quantum mechanical model was developed for an array of coupled particle plasma by Pino et. al \cite{Pino2014} whereby the model was used to predict the coupling strength between plasmon that surpasses or approaches the local dissipation, with quantum entanglement preserved in the collective modes of the array. In 2012, a method for creating maximally entangled states comprising two field quanta using a negative index metamaterial was proposed in \cite{Siomau2012}. In their scheme, two light fields of weak intensity are prepared in either polarization or coherent states and then caused them to interact with a composite medium near an interface between a negative-index metamaterial and a dielectric. Also, Song et. al \cite{Song2013} proposed a model for obtaining Bell-type entangled coherent states for quantum information processing using Cross-Kerr nonlinearity in a metamaterial. However, one of the challenges with plasmonic metamaterial is the loss. To reduce the loss usually encountered with plasmonic metamaterial, Moiseev et. al \cite{Moiseev2016} proposed a low-loss nonlinear negative index metamaterial.  In their scheme, two low-loss surface polariton that interact through the negative-index metamaterial retain their low-loss nature and yield large mutual $\pi$ phase shift. Also, it has been reported by Wuestner et. al \cite{Wuestner2010} that loss could be overcome with gain in a negative-index metamaterial. Yannopapas \cite{Yannopapas2015} in 2015 used a quantum correction model to study the effect of the electron tunnelling between two metallic surfaces in the optical response of a three-dimensional plasmonic metamaterial with subnanometer gaps. Of recent, Sowa and Zagoskin \cite{Sowa2019} propose a model for an exactly solvable quantum metamaterial when a quantum metamaterial interacting with an electromagnetic field is considered. In their work, they showed that the interaction of the quantum metamaterial with the electromagnetic field could be solved with the combination of the Haar transform and the Wigner-Weyl transform.

\section{Fabrication}
In this section, we briefly review the fabrication methods and techniques for QMMs. A QMM is fabricated in a similar way to other optical metamaterials. The fabrication methods are classified into four categories: direct write, pattern transfer, hybrid patterning lithography, and with the so called alternative techniques \cite{su2018}. However, the most commonly used one is the direct write method which consists of electron beam lithography (EBL), focused-ion beam, probe scanning, direct writing, and interference lithography.
\newline EBL is the process whereby a focused Gaussian particle beam or focused electron is used to create custom shapes on an electron-sensitive film (resist) covered surface. These patterns are created either by a bottom-up or top-down approach. Both approaches are considered standard in metasurfaces fabrication. The choice of the method utilised depends on how the quantum metamaterials are made. The bottom-down approach uses the deposition or evaporation of molecules or atoms alongside with the lift-off process to create nanostructures or patterns, whereas the top-down approach uses the etching process to create nanostructures. EBL is the most commonly used method in fabricating quantum metamaterials because of the high resolution, good quality and ease of use. Focused ion beam (FIB) is another direct write method which is commonly used  to fabricate very small structures such as nanostructures or integrated circuits. FIB is a sputtering process whereby ion beam is used to mills the sample surface to create patterns with nanometer precision. This method allows milling under direct visualization, but it is not suitable for large-area manufacturing due to high cost and low throughput. Some other limitations of FIB are the low aspect ratio, long milling time which could lead to spatial drift of the sample image and sometimes, the sample damage during milling and imaging. Interference lithography (IL) is a laser-based technique for creating irregular arrays patterns of nanostructures without the use of photomask or complex optical systems. The advantage of this technique is the high fabrication speed and large area. IL is not a good method to fabricate periodic nanostructures such as plasmonics metasurfaces. Direct laser writing (DLW) is a maskless laser-based technique that uses computer-controlled optics to project the required nano-patterns directly onto the photoresist by holding the mask in software. It is suitable for large area fabrication and has high precision, good uniformity and low fabrication cost. The disadvantage of this technique is that there is no batch process.
\paragraph{} Pattern Transfer Lithography (PTL) is a method developed to solve the problem of high throughput and large area during fabrication of the nanostructures. PTL include plasmonic lithography (PL), nano-imprint lithography (NIL) and self-assemble lithography (SAL). PL is developed to have deep subwavelength resolution beyond the diffraction limit, unlike the traditional photolitography that suffers from diffraction limit. The ability to create subwavelength resolution  and high throughput are the strength of this technique, but large area photomask remains a challenge to overcome. NIL is a technique for creating nanoscale patterns by mechanical deformation of imprint resist and subsequent process. NIL has a high resolution, low cost, and high throughput. However, the residual imprint layer is the main challenge of this technique. Self-assembly lithography is a method whereby nanostructures are created by self-assembly. This method enables large area fabrication, low cost and high throughput. The setback of this method is the lack of uniformity and limited patterns.
\paragraph{}Hybrid lithography is a method that combines the above mentioned lithography techniques to realize metasurfaces with more complex structures, such as those relevant to quantum metamaterials. Hybrid lithography includes microsphere projection lithography (MPL) and hole-mask colloid and off-normal deposition lithography (HMCODL). MPL is defined by local projection lithography, with a single microsphere acting as a lens.  It uses self-assembled arrays of silica spheres as colloidal microlenses whereby each microlens project an image of a distant macroscopic mask onto the coated substrate. MPL is developed for rapid prototyping of periodic and quasi-periodic metasurfaces. HMCODL is a new inexpensive technique that is based on colloidal self-assembly lithography patterning to make directional metasurfaces \cite{verre2016}. This technique is suitable for large area and making of tilted structures. However, since it combines the self-assembly technique, sample uniformity is still a challenge to overcome.
\paragraph{}An alternative technique is developed to cater for the limitation of the above mentioned methods. This technique is new, and it includes forward transfer lithography, ablation and dewetting lithography, and two photon lithography. Forward transfer lithography is a high throughput and versatile technique that is based on the laser direct write technique. It is a one-step technique for fabricating periodic and non-periodic nanostructures. However, a supporting transparent substrate is needed for this technique and no batch processes. Ablation and Dewetting is also a laser based technique where nano-materials are lifted off the surface of a substrate with a focusing laser. It is a single step, cost efficient, lithography free technique for large scale process. The last in this category is the two photon lithography (TPL) technique. TPL uses an ultrashort laser pulse to create nano-patterns for making 3-D nanostructures. The advantage of TPL includes good geometry control, cleanroom free, and scalable resolution. However, the limitation of photosensitive materials and low efficiency of two-photon absorption remains a challenge to overcome.


\section{Applications of quantum metamaterials to quantum information processing}

In this section, we review the applications of QMM to quantum information processing. QMM are an important optical material in quantum information processing due to their unique ways of controlling and manipulating electromagnetic waves. This has been exploited in a wide variety (various) of applications, including wavefront shaping \cite{li2019,Ding2015,chen2018}, polarisation control \cite{li2015,zhao2012}, performing quantum state engineering \cite{uriri2018}, quantum entanglement \cite{Asano2015}, Quantum algorithm \cite{zhang2018}, orbital angular momentum generation \cite{Bouchard2014}, on-chip quantum information processing \cite{szedlak2016}, and many more.  For clarity, we report in detail the applications of quantum metamaterials to quantum information processing in subsections A to H as follow:

\subsection{Quantum State Manipulation and Control}
Quantum metamaterials have opened up new ways of manipulating and controlling the amplitude, phase, and polarisation of light at the single photon level for quantum information processing. Initially, the fabrication of metamaterials aimed to provide new materials with unusual passive electromagnetic properties such as artificial magnetism \cite{Xiao2009} and negative refractive index \cite{Ivic2014}. However, this whole concept may have been pushed to another level with the introduction of quantum degrees of freedom. This was achieved by embedding highly controllable quantum systems, such as quantum dots, cold atoms, Josephson junctions, or molecules into the metamaterial itself \cite{Rakhmanov2008, Anlage2011, Pashkin2009, Shvetsov2013, Jung2014}. Due to the quantum nature of these newly added media, the metamaterial possesses many novel features and ways of controlling and manipulating electromagnetic waves. The newly added quantum media exhibit controllable quantum states that retain quantum coherence times much greater than the traversal times of an electromagnetic wave. Recently, quantum metamaterials for controlling \cite{Rakhmanov2008, Felbacq529, Song2018} and manipulating \cite{Ivic2014, Wang2012, Wang2018} single photons for quantum information processing have been studied. The fundamental works of Rakhmanov et. al \cite{Rakhmanov2008} and Wang et. al \cite{Wang2012} provided a new platform for using quantum metamaterials to control and manipulate light. In 2008, Rakhmanov et. al \cite{Rakhmanov2008}, used a quantum metamaterial to control the propagation of electromagnetic waves in a way that is not possible with a normal classical structures. Wang et. al \cite{Wang2012} obtained Hong-Ou-Mandel interference mediated by the magnetic plasmon waves in a three-dimensional metamaterial. Also, the control of quantum state has been reported in Josephson junction based controllable quantum two-level systems for quantum computing using superconducting qubits \cite{Ivic2014}. The same year, Bouchard et. al \cite{Bouchard2014} demonstrated optical spin-to-orbital angular momentum conversion in a quantum metamaterial or ultrathin metasurface with arbitary topological charges. Another important application of quantum metamaterial is found in their ability to control the polarization of single photons effectively. Polarization  is an important aspect of quantum communication since it  can  be used  to  transmit  single photons  and  make  sensitive  quantum measurements. The use of a plasmonic metamaterial in polarization control of single photons for quantum information processing has been very successful in recent years. For example, Ren et al. \cite{Ren2017} reported a novel optically reconfigurable plasmonic hybrid metasurface that enables polarization tuning at optical frequencies. Chen et. al \cite{chen2018} used a geometric metasurface based on chiral plasmonic stepped nanoapertures to demonstrate spin-controlled wavefront shaping whereby the transmission from a sub-array is transmitted and the other is blocked. 

\subsection{Quantum Entanglement}
Quantum entanglement is an important information resource in quantum information processing and one of the major applications where quantum metamaterials have been used extensively. The work of Siomau et. al \cite{Siomau2012} in 2012 is a motivation for using QMM for generating quantum entanglement whereby maximally entangled states were created using a negative-index metamaterial.  In 2013, Song et. al \cite{Song2013} theoretically designed a quantum metamaterial model for obtaining Bell-type entangled coherent states for quantum information processing. Thereafter, Pino \cite{Pino2014} in 2014, detected and stored entangled states in the collective modes of the array of a plasmonic metamaterial. Their work drove immense interest in the field, which later lead to more practical ways of generating quantum entanglement using a quantum metamaterial; for example, the use of a plasmonic metamaterial for the distillation of partially entangled Bell states \cite{Asano2015}.

\subsection{Single Photon Generation}
In quantum information processing, single photons are usually called qubits, which are the basic unit of quantum information. Qubits are the quantum analogue of a bit in classical computing. In quantum optics, single photons are mainly generated by using a non-linear crystal (BBO) through spontaneous parametric down conversion (SPDC) \cite{Thorn2004, Bronner2009, Pearsona2010}, a nitrogen vacancy defect in diamond \cite{Schroder2011}, and/or through solid state qubits (superconducting qubits) \cite{Kindel2016}, cold atoms \cite{Farrera2016, Ho2018}, and quantum dots \cite{Hanschke2018}. Some of the physical materials used for these processes are cumbersome, for example, the optics used in the SPDC process which is more common is not suitable for on-demand single photons. Therefore, the need for better ways of generating single photons in smaller material (of the order of nanometers) for quantum information processing is necessary. This is also where quantum metamaterials could provide/deliver a significant practical advantage. In 2016, Poddubny et. al \cite{Poddubny2016}, theoretically obtained single photons or photon-plasmon quantum states in a non-linear hyperbolic metamaterial with 70 $\%$ internal heralding quantum efficiency. In their work,  a pair of signal and idler single photons from a laser beam through spontaneous non-linear wave mixing inside metal-dielectric structures were reported.

\subsection{Quantum State Switching}
The ability to tune the frequency, phase, amplitude of light is essential to quantum information processing, most especially doing this in nanomaterials because of their size. “Quantum metamaterials have already/previously been used for quantum switchable related tasks \cite{Hutter2011, Macha2014, Felbacq529, Hierro2019}. Microwave band gap tuning was realized by Hutter \cite{Hutter2011} using one dimensional Josephson junction arrays with generalized unit cells. The gap was tuned in a wide frequency range when an external flux was applied. In 2014, Macha et. al \cite{Macha2014}  realized superconducting  qubits using quantum metamaterials to implement the AC-Zeeman shift of a resonant qubit. In 2015, Felbacq and Roussear \cite{Felbacq529} reported all-optical photonic band control in a quantum metamaterial, embedding quantum resonators within the nanorods. Their quantum metamaterial exhibits an all-optical switchable conduction band. Moreover, Shulga et. al \cite{Shulga2018} observed frequency tunable transparency in a microwaved transmission line using a quantum metamaterial composed of twin flux qubits while more recently, Hierro et. al \cite{Hierro2019} reported a phase transition in a semiconductor quantum metamaterial by tuning the quantum wells thickness and the barrier height. 

\subsection{Quantum State Engineering}
Quantum state engineering is the active control and manipulation of quantum states using quantum media. The ability to actively control and manipulate quantum states with quantum metamaterials may lead to engineering metamaterial-based quantum transistors and switches. Recently, we reported the active control of a plasmonic metamaterial for quantum state engineering \cite{uriri2018}.  In our work, we actively tuned the transmission of single photons through plasmonic metamaterials when heat from an external laser was applied. However, the sensitivity and overall performance of this type of device can be improved with the addition of some non-linear media.

\subsection{Quantum key Distribution}
Quantum key distribution (QKD) is based on the transfer of single photons from one point to the next with the aim to distribute a secure key that can be utilised for encryption. The single photons are hence referred to as the quantum carriers as they carry vital information for the key distribution process. QKD is based on the design and assemble of physical optical systems to ensure security of information, and practically has been shown to provide a future-proof technology for secure communications that require confidential information to remain secure for extended periods of time \cite{Bennet92, Ekert}. 
QKD has been implemented via optical mediums such as fibre \cite{gisin2004,curty2014,Zbinden2015,Frohlich2017, peev2009,wang2014,Zhang} and free-space which includes ground-to-ground \cite{ursin2007, Krenn} and satellite based communication \cite{Yin,Liao}. Pan et al is leading the efforts towards the development of a quantum network in free-space, which includes satellite based communication, and fibre based intercity links \cite{Yin,Zhang, Liao}. QKD has also been achieved underwater via submarines \cite{Cheng} and with the drive towards miniaturisation and low cost, the development of QKD technology is pushing the limits toward using nano-satellites known as CubeSat \cite{Ling}. Machine learning has been applied to QKD  to improve the quantum channel and enhance the security during transmission \cite{Ismail2019,Li:18,Lau,Jiang,Ou,Zeng}. Quantum Plasmonics, due to its versatility and ability to ultimately miniaturize photonic components for quantum optic, is a new niche area which will be able to enhance QKD technology \cite{Mortensen}.

\subsection{Quantum Algorithm}
Quantum database search is important to quantum computing. For example, the Grover search algorithm achieves the task of finding the target element in an unsorted database in time quadratically faster than the classical computer. A similar search algorithm was shown to be possible with a quantum metamaterial. Recently, Zhang et. al \cite{zhang2018} implemented a quantum search algorithm using quantum metamaterials. The quantum search algorithm is performed when an incoming wave that passes through the designed metamaterials and the profile of the beam wavefront is processed repeatedly as it propagates through the unit cells arranged periodically in the metamaterial. Search items are found with the incoming wave all focusing on the marked positions after $\approx$ $\sqrt{N}$ roundtrips.

\subsection{Orbital angular momentum}
Solutions to the Helmholtz equation produces an entire family of optical beams carrying Orbital Angular Momentum (OAM) \cite{Allen}. These optical beams have been applied to numerous fields such as optical manipulation \cite{Grier,Rubinszteindunlop,Dholakia}, quantum optics \cite{Dada,Padgett2010}, optical communications \cite{Sonja,Yan,Willner}, quantum communication \cite{Zeilinger2002,Krenn,Villoresi,Forbes} and many other multidisciplinary research areas \cite{Torner}. There are numerous experimental methods to produce these beams \cite{Allen, Beijerersbergen, Beijerersbergen94,Heckenberg,Bazhenov,Turnbull,Arlt,Ismail2012}. While Beijersbergen et al. \cite{Beijerersbergen, Beijerersbergen94} were the first to demonstrate the existence of OAM beams by a method known as mode conversion, easier approaches are now used which include the use diffractive optics and digital holography.  Metasurfaces can be appropriately designed to produced tunable OAM of surface plasmon polaritons, achieved by controlling both the orientation angle and spatial position of nano aperture array on an ultrathin gold film thus producing a field distribution of the surface waves  which can be engineered to comprise both spin dependent and independent OAM components \cite{Tan}. Furthermore, the interaction between optical vortices and plasmonic nanoslits produces vortex beams,  which as shown in \cite{Min} may be used to detect the nanos- tructure on metal films. This provides opportunities for OAM detection on-chip \cite{Xie2017}.

\section{Summary}

“In this work, we have briefly reviewed current progress on quantum metamaterials with applications to
quantum information processing. For some years metamaterials research has attracted much attention most in the classical regime, leading to applications ranging from electromagnetic cloak to super-resolution imaging. Still, the pursuit of how this unique artificial material "metamaterial" can be applied to quantum information processing is an open question, although ongoing research has yielded some positive results. In the quantum regime, applications such as wavefront engineering, performing quantum state engineering, polarisation control, orbital angular momentum generation and on-chip orbital angular momentum creation have all been
obtained. However, new methods that are cost effective, suitable for large fabrication with good aspect ratio are still necessary to develop. Plasmonics and superconducting metamaterials may be a promising candidate for on-chip high dimensional quantum communication.

\section*{Acknowledgment}
This work is based on research supported by the South African Research Chair Initiative of Department of Science and Technology and National Research Foundation as well as the Thuthuka grant. Opinions expressed and conclusions arrived at, are those of the authors and are not necessarily to be attributed to the NRF.
\newline
\newline
\noindent\textbf{Disclosures.} The authors declare no conflicts of interest.

\bigskip

\bibliography{ref}

\bibliographyfullrefs{sample}

\end{document}